# The roles of wettability and surface tension in droplet formation during inkjet printing

Bing He, Sucui Yang, Zhangrong Qin, Binghai Wen[*], and Chaoying Zhang[*]

Guangxi Key Lab of Multi-source Information Mining & Security, Guangxi Normal University Guilin 541004, China

Correspondence and requests for materials should be addressed to:
oceanwen@gxnu.edu.cn, zhangcy@gxnu.edu.cn

This paper describes a lattice Boltzmann-based binary fluid model for inkjet printing. In this model, a time-dependent driving force is applied to actuate the droplet ejection. As a result, the actuation can be accurately controlled by adjusting the intensity and duration of the positive and negative forces, as well as the idle time. The present model was verified by reproducing the actual single droplet ejection process captured by fast imaging. This model was subsequently used to investigate droplet formation in piezoelectric inkjet printing. It was determined that wettability of the nozzle inner wall and the surface tension of the ink are vital factors controlling the print quality and speed. Increasing the contact angle of the nozzle inner delays the droplet breakup time and reduces the droplet velocity. In contrast, higher surface tension values promote earlier droplet breakup and faster drop velocity. These results indicate that the hydrophilic modification of the nozzle inner wall and the choice of inks with high surface tensions will improve printing quality.

Over the past few decades, inkjet printing technology has been widely used in various emerging industrial applications, including fabricating flexible displays, lab-on-a-chip devices, fuel injection, cell printing, and drug delivery.[1-3] However, these applications come with several serious challenges regarding print quality, as well as the requirements for higher speed and accuracy conjunction with increasingly with small ink droplets. Moreover the printing quality depends directly on the ink droplet formation process. Many researchers devote themselves to studying this method by using experiments and numerical simulations.

Experiment is a straightforward way to investigate inkjet printing and receives much attention from researchers. Based on various technological advances in other fields, especially in the development of sensitive high-resolution cameras capable of capturing instantaneous droplet shapes, it is now possible to assess the effects of nozzle size, voltage signal, jetting speed, droplet shape and ink properties.[4-9] Various experiments have advanced our understanding of the droplet formation process. However, these studies have also been primarily limited to obtain global estimates and visualizations of exterior of the inkjet nozzle.[4,10,11] In contrast, numerical simulations can be employed to determine the fluid dynamics throughout the entire droplet formation process and provide insight into the parameter conditions.

The drop-on-demand (DOD) inkjet printing method, especially piezoelectric inkjet (PIJ) technology, is the most commonly used for modern industrial applications, because it is easily controlled by tuning the driving electrical signal and is compatible with various materials. Fromm[12] used a marker-and-cell (MAC) method to analyze the dynamics of droplet formation from a DOD nozzle, although this method suffered from the lack of accuracy that is inherent to the MAC technique. Badie and de Lange[13] applied the finite-element method to simulate DOD droplet formation, while Feng[14] carried out a series of simulations of droplet ejection based on the fluid volume method. Yu *et al.*[15] developed a coupled level set-projection approach to simulate the piezoelectric inkjet printing and analyzed the interface motion, droplet pinch off and satellite droplet formation. Xu and Basaran[16] simulated the formation of DOD droplets by means of the finite-element method and an inflow boundary condition, which corresponded to one full cycle of a sinusoidally varying inlet velocity. Although the above methods have generated reasonable estimates of the ejected droplet volume and velocity, these conventional Navier-Stokes-based calculations typically require significant computing power and cost. This is especially the case when numerical difficulties in the treatment of topological deformation of interface breaking and coalescing.[17,18] These issues severely limit the application of these methods to the study of inkjet printing, since inkjet systems include highly complicated topological variations at interface during droplet formation. The ability to numerical method to track the evolution of a free surface is critical, especially during breakage and coalescence of interfaces. In addition, these methods are based on continuum theory and need to solve the complex Navier-Stokes equation directly, which is also a complicated process. Therefore, simpler and more effective methods of simulating ink droplet formation would be highly desirable.

Inkjet printing quality is closely related to many of factors, such as the nozzle geometries, the nozzle materials, the ink properties, and the actuating conditions.[1,3] Two of the most important factors are the wettability of the nozzle and the surface tension of the ink, and these play important roles in the droplet formation process. During inkjet printing, the wettability of the nozzle has a significant effect on the formation, velocity and shape of the droplets. However, experimental studies face difficulties when attempting to study a wide range of nozzle wettability values, especially with respect to the wettability of the nozzle inner wall. Recently, some experimental and numerical studies have been presented focusing on the wettability of the nozzle tip and the nozzle plate surface.[1,19,20] In Electrohydrodynamic jet, some studies have paid special attention on the shape of the meniscus, which is affected by the wettability of the outer surface of the nozzle tip.[19,20] Lai et al.[21] investigated the effect of the dynamic contact angle during the ejection of a droplet. In contrast to PIJ, this work used a vibrational motion of the nozzle plate to generate the oscillating pressure difference. However, up to now, there have been few reports of the effects of the wettability of the nozzle inner wall. In addition, surfactants are routinely used in printing ink to control the breakup of droplets in traditional and emerging applications of inkjet printing.[3] A comprehensive understanding of surface tension effects on inkjet printing quality would be helpful.[3] Dong et al.[6] investigated the effects of two fluids with different surface tensions on DOD droplet formation. Suryo and Basaran[22] simulated the thermally driven droplet-formation method and observed surface tension effects. Yang et al.[23] exploited CFD software to explore the droplet ejection behavior of a Picojet printhead and determined the effects of the physical properties of the ink. However, the effects of the ink viscosity and surface tension cannot be distinguished using these methods, even though it would be helpful to separately study the effects of the surface tension on PIJ print qualities.

Working at the mesoscopic level, the lattice Boltzmann method (LBM) is particularly suitable for the analysis of complex fluid systems involving interfacial dynamics and phase transition.[17,18,24-27] Using this model, the evolution of interface can be readily obtained without front-capturing and front-tracking treatments and this is a significant advantage when studying inkjet printing. Our aim in the present work was to apply a multiphase LBM to create an easily implementing, highly efficient and robust model for studying the droplet formation during the DOD printing process. Through comparison with experimental results, we were able to verify the capability and efficiency of this model. The model was subsequently applied to systematically investigate the impacts of the wetting properties of the nozzle inner wall and the surface tension of the ink on inkjet print quality. The droplet formation process was analyzed in-depth, including assessments of droplet shapes, pinch-off time, and velocity variations.

## Results

**Actuation modeling**

The actuation wave is an important parameter in the adjustment of the droplet formation. In PIJ printing, the ejection process is actuated by a driving voltage that induces a pressure wave inside the nozzle. Bogy et al.[28] detailed the pressure fluctuations induced by a single positive voltage wave (as so-called "on-off" pulse).

As the voltage rises, a positive pressure wave appears inside the nozzle and generates the pushing force that results in ink ejection. Then, after a voltage hold time, the voltage falls and a negative pressure wave appears inside the nozzle and produces the pulling force that leads to the separation of the ejected droplet from the nozzle. To mimic the pushing and pulling force induced by the pressure fluctuations, we employed the time-dependent driving force pattern shown in Fig.1. After lots of trial and error, a full time period is set to a positive force pulse of 5 μs duration, a negative force pulse of 3 μs duration and an idle duration of 8 μs. By changing the intensities and the hold times of the positive, negative forces and the idle time, the desired actuating effects could be readily obtained.

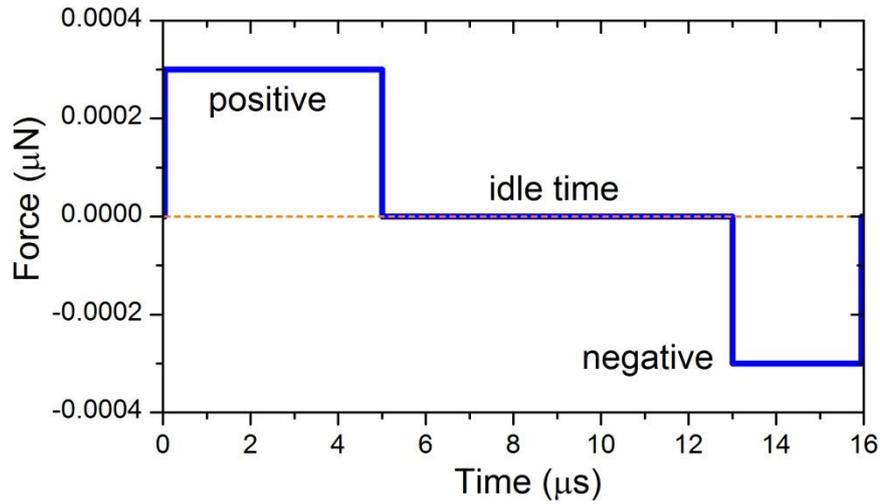

FIG. 1. Time-dependent driving force applied in the present simulations.

**Reproducing the experiment**

Numerical simulations must be consistent to the real world. With this foundation, it is able to capture more information than an experiment. Recently, van der Bos *et al.*[4] obtained the images of the formation of small droplets at high speeds on the nanosecond level. They employed a silicone oil with density of 930 kg/m$^3$, viscosity of 9.3 mPa s and surface tension of 20.2 mN/m in conjunction with a nozzle radius, R=15 μm. We reproduced this experimental work in our simulations as means of validating the present model. In our lattice Boltzmann model, the computational domain is a rectangular with the length 800 and the width 200 lattice units. The height and radius of the nozzle is 120 and 40 lattice units, respectively. The density of liquid phase is $\rho_l$=1000, the density of gas phase is $\rho_g$=1, and the width of the interface layer is W=3. The relaxation parameters take $\tau_n = 0.85$ and $\tau_\phi = 0.8$. The kinematic viscosity is $v_l = 0.17$ and the surface tension is σ=1.08. The periodic boundary condition is used on both of the left and right sides, while the fully developed boundary condition is used on the top and bottom.[17] The wetting boundary condition is used on the nozzle inner wall and the contact angle of the nozzle inner wall is 90°.[29]

The resulting successive images generated at time intervals of 10 μs are presented in Fig. 2. These images allow ready inspection of the droplet ejection process. At the first 10 μs, the fluid begins to flow rapidly out of the nozzle orifice, as a result of the positive force pulse. From 10 μs to 50 μs, the ejecting fluid is increasingly stretched and forms a rounded main drop with an attached filament. The filament breaks up between 50 μs to 60 μs and the tail end becomes rounded beginning at 70μs, after which the tail end of the filament recoils and speeds up as a result of surface tension, such that the separated droplet contracts. The contracting filament pulls on the head and the entire mass shrinks into a single droplet up to 130 μs. Finally, a steady droplet with a near-spherical shape is formed, having a velocity of approximately 1.5 m/s. From Fig. 3, it is evident that the velocities of both the droplet head and tail undergo perturbations. The internal oscillations arise from the initial violent jetting of the ink. The magnitude of these perturbations is gradually reduced over time due to surface tension effect. Similar perturbations have also been observed in the experiments, in which van der Bos *et al*. used momentum to depict the movement and merging of the droplet head and tail.[4] The numerical results obtained in the present work agree very well with the experimental images.[4,5] Hence, the present LBM is evidently a useful tool for the investigation of droplet formation in inkjet printing.

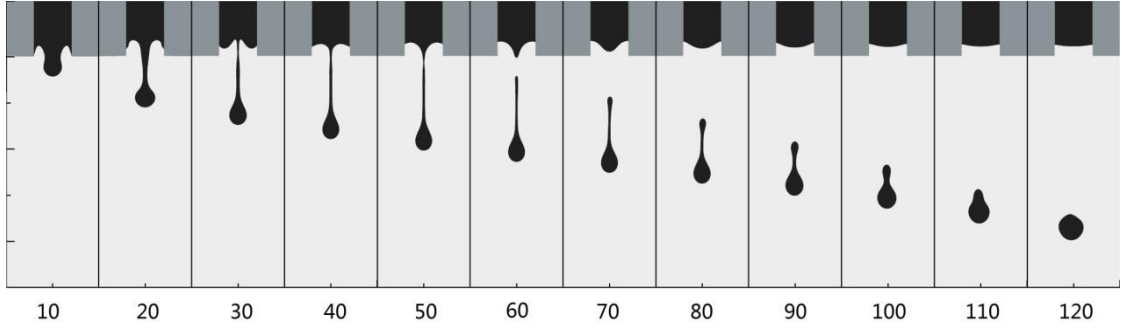

FIG. 2. Droplet formation images acquired at 10μs intervals.

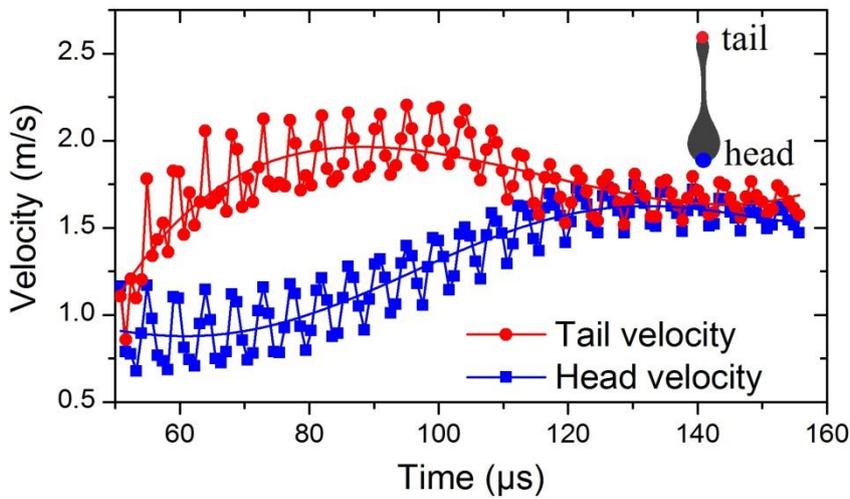

FIG. 3. Velocity profiles of the drop tail and head after the droplet breakup.

## Discussion

Based on the work described above, we separately investigated two inkjet printing parameters, wettability of the nozzle inner wall and surface tension of the ink, both of which play important roles in controlling the print quality and speed. Initially, using the same silicone oil ink, we adjusted the property of the nozzle inner wall to obtain varying degrees of surface wettability. These variations were reflected in changes in the contact angle from 0° to 180°. Because the inner wall of the nozzle directly touches the ink, its wettability can be expected to significantly affect the adhesive force.

Fig. 4 shows the contours of the flow field at the moment of droplet breakup. These images demonstrate that, up to the point of breakup, a more hydrophobic nozzle allows the droplet head to fly further and the droplet filament to extend over a greater distance. The hydrophilic walls also show stronger adhesion to the ink such that, at the contact angle less than 60°, the meniscus is concave and the breakup position is inside the nozzle. As the hydrophobicity of the wall increases, the meniscus becomes increasingly convex and the breakup position is out of the nozzle. Fig. 5 demonstrates that the wall wettability significantly influences both the breakup time and the droplet velocity. The hydrophobic nozzle obviously postpones the droplet breakup, such that the ink droplet in Fig. 4 generates a long, thin filament. The long filament is a main factor increasing the probability of satellite droplet forming, which will reduce the print quality.[1,6] Since a long filament on the ink droplet acts like a spring between the drop and the nozzle, it tends to slow the jetting drop, even when the hydrophobic wall generates a weak adhesive force. A slow ink droplet also inevitably degrades the efficiency of the inkjet printing. In contrast, when the nozzle is hydrophilic, the droplet breaks up earlier, such that the ink droplet in Fig. 4 generates a shorter, thicker filament. Comparing the data at contact angle 30° to 150°, the droplet breakup is advanced by 25 μs while the velocity increases approximately 16%. Therefore, a hydrophilic inner wall is a beneficial design feature that improves both the print quality and speed. Fig. 5 also demonstrates that a contact angle less than 30° leads to diminishing improvements, suggesting that it is not necessary to use an extremely hydrophilic nozzle.

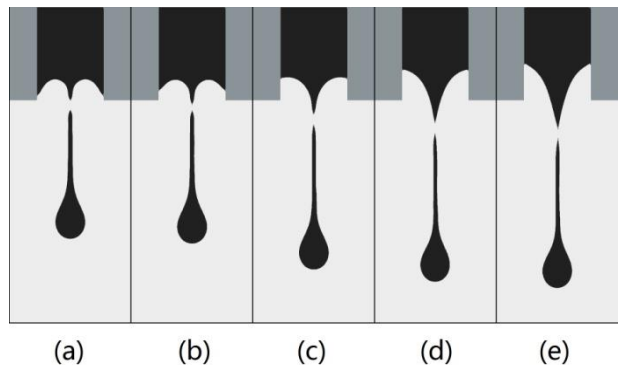

*FIG. 4.* Ink droplet images at the moment of breakup at nozzle inner wall contact angles and the corresponding breakup times of (a) θ=30°, breakup time 37.14 μs; (b) θ=60°, breakup time 38.78 μs; (c) θ=90°, breakup time 51.89 μs ;(d) θ=120°, breakup

time 58.45 μs; (e) θ=150°, breakup time 61.73 μs.

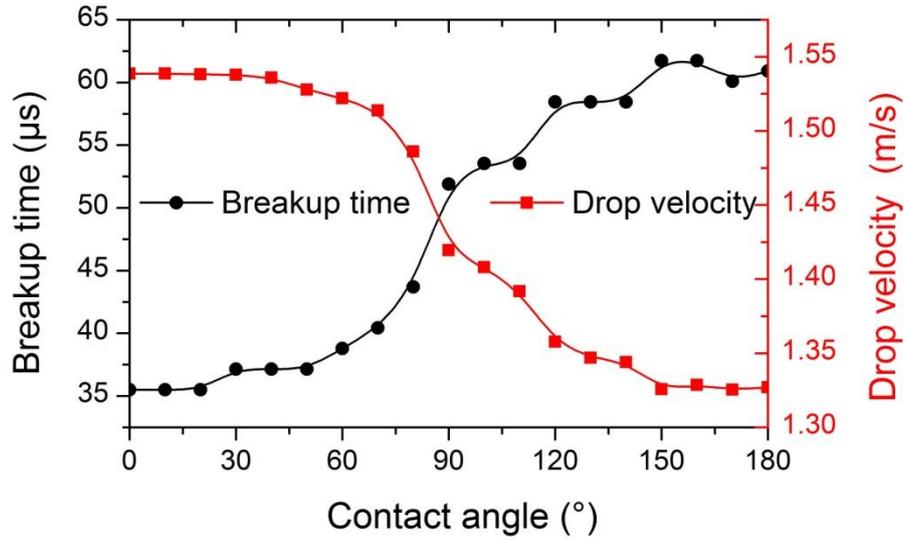

FIG.5. Variations in the breakup time and droplet velocity with the changes in wettability of the nozzle inner wall.

In subsequent trials, maintaining the same nozzle wettability and actuation force, we adjusted the surface tension of the ink in gradually increments from 20 to 85 mN/m. The surface tension will significantly affect the formation and evolution of the ink droplet, since surface tension tends to pinch off the droplet tail and shrink the ink droplet toward a spherical shape. This tendency is obvious in the images in Fig. 6. As the droplet breakup progresses, the droplet with a higher surface tension shrinks more rapidly and exhibits a far shorter tail. Therefore, a low surface tension ink generates droplets that tend produce satellite droplets. Fig. 7 further summarizes the relationship between the breakup time, droplet velocity and surface tension. With increasing surface tension, the breakup occurs sooner; a surface tension of 85 mN/m results in a breakup that occurs 18 μs before that with surface tension of 20 mN/m. These simulation results are in good agreement with results reported in Ref. 6. In addition, Fig.7 shows that the surface tension increases the droplet velocity in an almost linear manner. Comparing the results at surface tensions of 85 and 20 mN/m, the ink droplet is accelerated by about 56%. Hence, a high surface tension ink can help to suppress the formation of satellite droplets and increase the printing velocity. As a consequence, both print quality and efficiency are improved when using an ink with a high surface tension. Based on these data, a surfactant should be used judiciously after careful consideration.

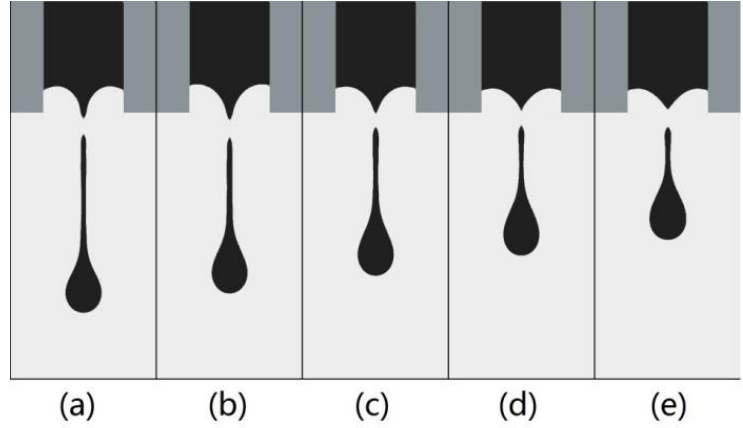

FIG. 6. Ink droplet images at the moment of breakup at surface tensions and corresponding breakup times of (a) σ=30 mN/m, breakup time 51.89 μs; (b) σ=40 mN/m, breakup time 48.61 μs; (c) σ=50 mN/m, breakup time 38.78 μs; (d) σ=70 mN/m, breakup time 33.68 μs; (e) σ=80 mN/m, breakup time 33.68 μs.

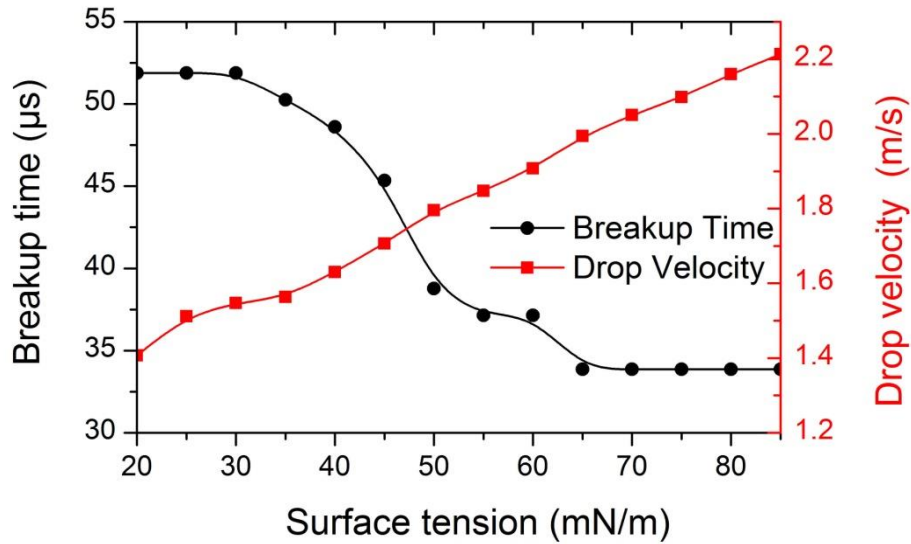

FIG. 7. Breakup time and drop velocity changed by the surface tension of the ink.

In summary, this work involved a numerical investigation of droplet formation during inkjet printing, using a LBM-based binary fluid model. Pressure fluctuations inside the nozzle were simulated by employing a time-dependent driving force to actuate the droplet ejection. This approach allowed the model to readily modify the droplet actuation by adjusting the intensity and duration of the positive and negative forces as well as the idle. This model was used to examine the single droplet formation during PIJ printing. A comparison of the numerical results with the experimental observations[4] found good agreements, demonstrating the ability of the present model to simulate the droplets formation in a PIJ device.

The model was applied to study of the roles of the nozzle wettability and the ink surface tension. A series of simulations were performed and the results were analyzed with respect to three important properties: the filament length, the breakup time and droplet velocity. As the contact angle of the nozzle inner wall was decreased from 150 ° to 30 °, the breakup time was decreased by 28 μs while the droplet achieved a higher speed. As well, increases in the ink surface tension from 20 to 85 mN/m moved the droplet breakup forward by 18 μs and increased the droplet velocity by approximately 56%. Both a hydrophilic nozzle inner wall and a high ink surface tension were found to shorten the droplet filament and avoid the formation of satellite droplet. These results indicate that the hydrophilic modification of the nozzle inner wall and the choice of inks having a high surface tension will improve print quality and performance.

## Methods

Basing on LBM, we select a free energy model with a high density ratio to simulate the gas/ink binary fluid system, in which the typical density ration as high as 1000.[30] The fluid flow is described by the Navier-Stokes equations

$$\frac{\partial n}{\partial t} + \nabla \cdot (n\mathbf{u}) = 0 \text{ and} \tag{1}$$

$$\frac{\partial (n\mathbf{u})}{\partial t} + \nabla \cdot (n\mathbf{u}\mathbf{u}) = -\nabla \cdot \vec{P} + \mu \nabla^2 \mathbf{u} + \mathbf{F}_b, \tag{2}$$

where n is the average density defined as $n = (\rho_A + \rho_B)/2$ ($\rho_A$ and $\rho_B$ being the densities of the two fluids, respectively), $\tilde{P}$ is the pressure tensor and $\boldsymbol{F}_b$ is the body force. They can be recovered by a lattice Boltzmann equation

$$f_i(\mathbf{x}+\mathbf{c}_i\delta t, t+\delta t) = f_i(\mathbf{x},t) + \Omega_i, \tag{3}$$

with the collision operator

$$\Omega_i = \frac{1}{\tau_n}[f_i^{(0)}(\mathbf{x},t) - f_i(\mathbf{x},t)] + (1-\frac{1}{2\tau_n})\frac{w_i}{c_s^2}[(\mathbf{c}_i-\mathbf{u}) + \frac{(\mathbf{c}_i\cdot\mathbf{u})}{c_s^2}\mathbf{c}_i](\mu_\phi\nabla\phi + \mathbf{F}_b)\delta t, \tag{4}$$

where $\delta t$ is the time step, $c_i$ is the lattice velocity, and $f_i^{(0)}$ is the equilibrium distribution function. For a two-dimension nine-velocity model, the equilibrium distribution function is read as

$$f_i^{(0)} = w_i A_i + w_i n(3c_{i\alpha}u_\alpha - \frac{3}{2}u^2 + \frac{9}{2}c_{i\alpha}c_{i\beta}u_\alpha u_\beta) \quad (i=0, 1\ldots, 8), \tag{5}$$

where the coefficients are taken as

$$\begin{cases} A_0 = \frac{9}{4}n - \frac{15(\phi\mu_\phi + \frac{1}{3}n)}{4}, & A_i\big|_{i=1,\cdots,8} = 3(\phi\mu_\phi + \frac{1}{3}n), \\ w_0 = \frac{4}{9}, & w_i\big|_{i=1,\cdots,4} = \frac{1}{9}, \quad w_i\big|_{i=5,\cdots,8} = \frac{1}{36} \end{cases} \tag{6}$$

The phase interface is captured by the well-known Cahn-Hilliard equation[31]

$$\frac{\partial \phi}{\partial t} + \nabla \cdot (\phi\mathbf{u}) = \theta_M \nabla^2 \mu_\phi, \tag{7}$$

where $\phi$ is the order parameter defined as $\phi = (\rho_A - \rho_B)/2$, $\mu_\phi$ is the chemical potential and $\theta_M$ is the mobility coefficient. It can be recovered by a modified lattice Boltzmann equation[30]

$$\begin{aligned} g_i(\boldsymbol{x}+\boldsymbol{c}_i\delta, t+\delta) \\ = g_i(\boldsymbol{x},t) + (1-q)[g_i(\boldsymbol{x}+\boldsymbol{c}_i\delta,t) - g_i(\boldsymbol{x},t)] + \frac{1}{\tau_\phi}[g_i^{(0)}(\boldsymbol{x},t) - g_i(\boldsymbol{x},t)] \end{aligned} \tag{8}$$

where $g_i$ is the order parameter distribution function, $\tau_\phi$ is the relaxation parameter, $q = 1/(\tau_\phi + 0.5)$ and $g_i^{(0)}$ is the equilibrium distribution function. A two-dimension five-velocity model is adopted to evolve Eq. (8) and the equilibrium distribution function is read as

$$g_i^{(0)} = A_i + B_i\phi + C_i\phi \mathbf{c}_i \cdot \mathbf{u} \qquad (i=0, 1\ldots, 4). \tag{9}$$

The coefficients are taken as

$$\begin{cases} B_1 = 1, \quad B_i = 0 \quad (i \neq 1), \\ C_i = \dfrac{1}{2q}, \\ A_1 = -2\Gamma\mu_\phi, \quad A_i = \dfrac{1}{2}\Gamma\mu_\phi \quad (i \neq 1) \end{cases} \tag{10}$$

1	Wijshoff, H. The dynamics of the piezo inkjet printhead operation. *Phys.Rep.* **491**, 77-177, (2010).
2	Singh, M., Haverinen, H. M., Dhagat, P. & Jabbour, G. E. Inkjet Printing-Process and Its Applications. *Adv.Mater.* **22**, 673-685, (2010).
3	Basaran, O. A., Gao, H. & P.P.Bhat. Nonstandard Ink Jets. *Annu. Rev. Fluid Mech.* **45**, 85-113, (2013).
4	Bos, A. v. d. *et al.* Velocity Profile inside Piezoacoustic Inkjet Droplets in Flight: Comparison between Experiment and Numerical Simulation. *Phys. Rev. Applied* **1**, 014004, (2014).
5	Physics: Fast imaging captures falling droplets. *Nature*. **507(7491)**, 142, (2014).
6	Dong, H. & Carr, W. W. An experimental study of drop-on-demand drop formation. *Phys. Fluids*. **18**, 072102, (2006).
7	Kwon, K. S. Speed measurement of ink droplet by using edge detection techniques. *Measurement* **42**, 44-50, (2009).
8	Castrejón-Pita, J. R., Morrison, N. F., Harlen, O. G., Martin, G. D. & Hutchings, I. M. Experiments and Lagrangian simulations on the formation of droplets in drop-on-demand mode. *Phys. Rev. E*. **83**, 036306, (2011).
9	Castrejón-Pita, A. A., Castrejón-Pita, J. R. & Hutchings, I. M. Breakup of Liquid Filaments. *PRL*. **108**, 074506, (2012).
10	Kim, C. S., Park, S. J., Sim, W., Kim, Y. J. & Yoo, Y. Modeling and characterization of an industrial inkjet head for micro-patterning on printed circuit boards. *Computers & Fluids*. **38**, 602–612, (2009).
11	Tan, H., Torniainen, E., Markel, D. P. & Browning, R. N. K. Numerical simulation of droplet ejection of thermal inkjet printheads. *Int. J. Numer. Meth. Fluids*. **77**, 544–570, (2015).
12	J.E.Fromm. in *Proc. 2nd Int. Colloq. Drops Bubbles,Monterey,California,Nov.19-21,1981* (ed Dennis H. Le Croissette) 322-333 (NASA JPL Publication 1981).
13	Badie, R. & Lange, D. F. d. Mechanism of drop constriction in a drop-on-demand inkjet system. *Proc. R. Soc. Lond. A*. **453**, 2573-2581, (1997).
14	Feng, J. Q. A General Fluid Dynamic Analysis of Drop Ejection in Drop-on-Demand Ink Jet Devices. *J. Imaging Sci. Technol.* **46**, 398-408, (2002).
15	Yu, J. D., Sakai, S. & Sethian, J. A coupled quadrilateral grid level set projection method applied to ink jet simulation. *J. Comput. Phys.* **206**, 227–251, (2005).

16  Xu, Q. & Basaran, O. A. Computational analysis of drop-on-demand drop formation. *Phys. Fluids.* **19**, 102111, ( 2007).

17  Aidun, C. K. & Clausen, J. R. Lattice-Boltzmann Method for Complex Flows. *Annu. Rev. Fluid Mech.* **42**, 439-472, (2010).

18  Chen, S. Y. & Doolen, G. D. Lattice Boltzmann method for fluid flows. *Annu. Rev. Fluid Mech.* **30**, 329-364, (1998).

19  Choi, K. H. *et al.* Development and ejection behavior of different material-based electrostatic ink-jet heads. *Int. J. Adv Manuf Technol.* **48**, 165-173, (2010).

20  Kim, Y. J. *et al.* Comparative Study on Ejection Phenomena of Droplets from Electro-Hydrodynamic Jet by Hydrophobic and Hydrophilic Coatings of Nozzles. *Jpn. J. Appl. Phys.* **49**, 060217, (2010).

21  Lai, J. M., Huang, C. Y., Chen, C. H., Linliu, K. & Lin, J. D. Influence of liquid hydrophobicity and nozzle passage curvature on microfluidic dynamics in a drop ejection process. *J. Micromech.Microeng.* **20**, 015033, (2010).

22  Suryo, R. & Basaran, O. A. Dripping of a liquid from a tube in the absence of gravity. *PRL.* **96**, 034504, (2006).

23  Yang, A. S., Yang, J. C. & Hong, M. C. Droplet ejection study of a Picojet printhead. *J. Micromech.Microeng.* **16**, 180-188, (2006).

24  Ladd, A. J. C. & Verberg, R. Lattice-Boltzmann simulations of particle-fluid suspensions. *J. Stat. Phys.* **104**, 1191-1251, (2001).

25  Zhang, J. Lattice Boltzmann method for microfluidics: models and applications. *Microfluid Nanofluid.* **10**, 1-28, (2011).

26  Wen, B., Qin, Z., Zhang, C. & Fang, H. Thermodynamic-Consistent Lattice Boltzmann Model for Nonideal Fluids. *Europhys. Lett.* **112**, 44002, (2015).

27  Wen, B., Zhang, C., Tu, Y., Wang, C. & Fang, H. Galilean invariant fluid–solid interfacial dynamics in lattice Boltzmann simulations. *J. Comput. Phys.* **266**, 161-170, (2014).

28  Bogy, D. B. & E.Talke, F. Experimental and theoretical study of wave propagation phenomena in drop-on-demand ink jet devices. *IBM J.Res.Develop.* **28**, 314-321, (1984).

29  Yan, Y. Y. & Zu, Y. Q. A lattice Boltzmann method for incompressible two-phase flows on partial wetting surface with large density ratio. *J. Comput. Phys.* **227**, 763–775, (2007).

30  Zheng, H. W., Shu, C. & Chew, Y. T. A lattice Boltzmann model for multiphase flows with large density ratio. *J. Comput. Phys.* **218**, 353–371, (2006).

31  Cahn, J. W. & Hilliard, J. E. Free energy of a nonuniform system. I. Interfacial energy. *J. Chem. Phys.* **28**, 258-267, (1958).

## Acknowledgements

This work was supported by the National Natural Science Foundation of China (Grant Nos. 11362003, 11462003, 11162002), Guangxi Natural Science Foundation (Grant No. 2014GXNSFAA118018), Guangxi Science and Technology Foundation of College and University (Grant No. KY2015ZD017), Guangxi Promoting Young and Middle-aged Teachers' Basic Ability Project (Grant No. KY2016YB063), Guangxi "Bagui Scholar" Teams for Innovation and Research Project, Guangxi Collaborative Innovation Center of Multi-source Information Integration and Intelligent Processing.

## Author contributions statement

B.H.W. and .C.Y.Z contributed most of the ideas. B.H.W., C.Y.Z., B.H. and S.C.Y. designed simulations. B.H. and S.C.Y. performed most of the numerical simulations. B.H.W., C.Y.Z., B.H. and Z.R.Q carried out theoretical analysis. B.H.W., C.Y.Z. and B.H. wrote the paper. All authors discussed the results and commented on the manuscript.

## Additional information

Competing financial interests: The authors declare no competing financial interests